\documentclass[prd, aps, superscriptaddress, preprintnumbers, twocolumn, floatfix, nofootinbib]{revtex4}
\pdfoutput=1

\usepackage{amsfonts}
\usepackage{amsmath}
\usepackage{amssymb}
\usepackage{bm}
\usepackage{dcolumn}
\usepackage{graphicx}   
\usepackage[latin1]{inputenc}
\usepackage{latexsym}
\usepackage{rotating}
\usepackage{hyperref}
\usepackage{graphicx}
\usepackage{color}

\newcommand\be{\begin{equation}}
\newcommand\ba{\begin{eqnarray}}
\newcommand\ee{\end{equation}}
\newcommand\ea{\end{eqnarray}}

\begin{document}

\title{Cosmic Strings from Thermal Inflation}

\author{Robert Brandenberger}
\email{rhb@physics.mcgill.ca}
\affiliation{Department of Physics, McGill University, Montr\'{e}al, QC, H3A 2T8, Canada}

\author{Aline Favero}
\email{aline.favero@mail.mcgill.ca}
\affiliation{Department of Physics, McGill University, Montr\'{e}al, QC, H3A 2T8, Canada}

\date{\today}

\begin{abstract}

Thermal inflation was proposed as a mechanism to dilute the density of cosmological moduli. Thermal inflation is driven by a complex scalar field possessing a large vacuum expectation value and a very flat potential, called a `flaton'.  Such a model admits cosmic string solutions, and a network of such strings will inevitably form in the symmetry breaking phase transition at the end of the period of thermal inflation. We discuss the differences of these strings compared to the strings which form in the Abelian Higgs model.  Specifically, we find that the upper bound on the symmetry breaking scale is parametrically lower than in the case of Abelian Higgs strings, and that the lower cutoff on the string loop distribution is determined by cusp annihilation rather than by gravitational radiation (for the value of the transition temperature proposed in the original work on thermal inflation).

\end{abstract}

\maketitle

\section{Introduction} 
\label{sec:intro}

Thermal inflation (not to be confused with {\it warm inflation} \cite{warm}) was proposed \cite{Lyth96} as a mechanism to dilute the number density of unwanted moduli quanta.  Moduli fields are predicted in many models beyond the particle physics Standard Model.  In such models, moduli quanta are often left over from the Big Bang period, produced at the end of a phase of primordial inflation, or generated during a compactification phase transition. The proposal of \cite{Lyth96} is to dilute the number density of moduli by invoking a period of inflation. This period should be sufficiently long to dilute the moduli, but short enough not to redshift the fluctuations which are generated in the primordial universe.

Thermal inflation is generated by adding a new matter sector involving a complex scalar field \footnote{A real scalar field with a symmetry breaking potential would lead to domain walls, a cosmological disaster \cite{DW}.} $\phi$ with a symmetry breaking potential $V(\phi)$,  and gauging the resulting $U(1)$ symmetry (this gauging is done since there is evidence that global symmetries are inconsistent with quantum gravity \cite{noglobal}).  Thermal inflation is assumed to occur in the radiation phase of cosmology. Thermal effects are assumed to trap $\phi$ at the symmetric point $\phi = 0$. At a temperature $T_i$ given by
\be
V(0) \, = \, T_i^4 \, ,
\ee
the potential energy of $\phi$ begins to dominate and inflation begins.  The coupling of $\phi$ to the thermal bath generates a finite temperature contribution to the effective potential
\be
V_T(\phi) \, = \, V(\phi) + g T^2 \phi^2 \, ,
\ee
where $g$ is the coupling constant describing the interactions between $\phi$ and the thermal bath.  At a temperature $T_c$ when the positive contribution to the curvature of the potential at $\phi = 0$ equals the absolute value of the (negative) curvature coming from the bare potential $V(\phi)$,  symmetry breaking sets in, $\phi$ rolls to the bottom of its potential and thermal inflation ends.  In order to generate the required hierarchy between $T_c$ and $T_i$, it was assumed that the bare potential $V(\phi)$ contains no quartic term. The hierarchy between $T_i$ and $T_c$ determines the number $N$ of e-foldings of thermal inflation
\be
\frac{T_c}{T_i} \, = \, e^{-N} \, .
\ee

Since the vacuum manifold of $\phi$ is $S^1$, cosmic string defects inevitably form in the phase transition which ends thermal inflation \cite{Kibble} (see \cite{CSrevs} for reviews of the role of cosmic strings in cosmology).  As the scale of symmetry breaking for thermal inflation is assumed to be of the order $m_0 \sim 10^2 - 10^3 {\rm{GeV}}$ one might - based on intuition from Abelian Higgs strings \cite{NO} - have expected the signatures of these strings to be negligible. As we show here,  thermal inflation strings have different properties compared to strings formed in the standard Abelian Higgs model. For the same value of the symmetry breaking temperature, thermal inflation strings have a parametrically larger mass per unit length than regular strings. Comparing strings with the same mass per unit length $\mu$, thermal inflation strings have a parametrically greater width than regular strings. These differences affect the distribution of string loops, and hence the observational consequences of the strings (see e.g. \cite{RHBrev} for a short review) need to be revisited.

In this article we use natural units in which the speed of light, Planck's constant and Boltzmann's constant are set to 1.  The Planck mass is denoted by $m_{pl}$. The mass per unit length of a string will be denoted by $\mu$, or often in terms of the dimensionless quantity $G \mu$, where G is Newton's gravitational constant.

\section{Thermal Inflation Strings}

The thermal inflation model \cite{Lyth96} assumes a potential given by
\be
V\left(\phi\right) \, = \, V_{0} - m_{0}^{2} \vert\phi\vert^{2} + \sum_{n=1}^{\infty} \frac{\lambda_{n} \vert\phi\vert^{2n+4}}{{m_{pl}}^{2n}} \, ,
\ee
where $V_0$ is tuned such that the potential energy in the vacuum manifold vanishes. The $\lambda_n$ are dimensionless coupling constants, and the mass scale $m_0$ determines the (negative) curvature of the potential at the origin. We shall consider a simplified potential containing only the $n = 1$ term (the contributions from the terms with $n > 1$ are Planck suppressed for the questions we are asking, i.e. those involving small field values) \footnote{The absence of a quartic term in the potential is crucial if we are to generate a hierarchy between $T_i$ and $T_c$.}. Thus, we consider the potential
\be \label{pot}
V(\phi) \, = \, V_0 - m_0^2 \vert\phi\vert^2 + \lambda \vert\phi\vert^6  m_{pl}^{-2} \, .
\ee
This potential is to be compared with the potential for the Abelian Higgs model which is
\be
V_{AH} (\phi) \, = \, \frac{\lambda_{AH}}{4} \bigl(\vert\phi\vert^2 - \eta^2 \bigr)^2 \, ,
\ee
where $\eta$ is the value of $\vert\phi\vert$ in the vacuum manifold, and $\lambda_{AH}$ is a dimensionless coupling constant.
 
 As we will see below, the hierarchy between $T_i$ and $T_c$ increases as $m_0$ decreases. To obtain an e-folding number $N \sim 10$ of thermal inflation a value of $m_0 \sim 10^2 - 10^3 {\rm{GeV}}$ was suggested \cite{Lyth96}.
 
 From (\ref{pot}) it immediately follows that the value $\eta$ of $\vert\phi\vert$ which minimizes the potential is given by
 \be \label{etavalue}
 \eta^2 \, = \, \bigl(\frac{1}{3} \bigr)^{1/2} \lambda^{-1/2} \frac{m_{pl}}{m_0} m_0^2
 \ee
 which is parametrically larger by a factor of $m_{pl} / m_0$ than what is obtained for an Abelian Higgs string given the same value of the curvature of the potential at $\phi = 0$. For the value of $m_0$ indicated above,  $\eta$ is of the order of $10^{10} {\rm{GeV}}$ and not $10^2 {\rm{GeV}}$ is it would be for an Abelian Higgs string with the same value of $m_0$.
 
 Demanding that the potential vanishes for $\vert\phi\vert = \eta$ yields
 \be
 V_0 \, = \, \frac{2}{3} \bigl( \frac{1}{3}\bigr)^{1/2} \lambda^{-1/2} m_0^3 m_{pl} 
 \ee
 which is also parametrically larger by a factor of $m_{pl} / m_0$ compared to the corresponding result for the Abelian Higgs string (taking coupling constants to be of the order $1$). This leads to the fact that the temperature $T_i$ corresponding to the onset of thermal inflation is parametrically larger than what one might have guessed from Abelian Higgs string intuition
 \be \label{onset}
 T_i \, \sim \, \lambda^{-1/8} \bigl( \frac{m_{pl}}{m_0} \bigr)^{1/4} m_0 \, .
 \ee
 On the other hand, the temperature at which the symmetry breaking phase transition takes place is given by
 \be \label{crit}
 T_c \, \sim \, m_0 \, ,
 \ee
 setting $g$ to be of the order $1$. Comparing (\ref{onset}) and (\ref{crit}) we see that it is precisely the enhancement factor discussed above which allows for a period of thermal inflation to take place.
 
 Let us now compare the width of a thermal inflation string with that of an Abelian Higgs string for the same value of the phase transition temperature $T_c$. The width is determined by minimizing the sum of the potential and gradient energy terms.  Increasing the width of the string costs potential energy while decreasing the width leads to an increase of the gradient energy.  For a straight string centered at $r = 0$ (in polar coordinates), the field configuration of a string with winding number $1$ can be written as
 \be
 \phi(r, \theta) \, = \, f(r) \eta e^{i\theta} \, ,
 \ee
 where the profile function $f(r)$ increases from $f(0) = 0$ to $f(r) = 1$ for $r > w$. The potential energy $\mu_p$ per unit length of the string can hence be estimated to be
 \be \label{tension}
 \mu_p(w) \, \sim \, \pi w^2 V_0 \, .
 \ee
 The scalar field angular gradient energy for a local string is cancelled by the gauge fields beyond a radius $r_A$ which is set by the gauge field mass.  For $r < w$ the angular gradient energy decays since $f(r)$ decays as $r$ decreases. Hence,  the mass per unit length $\mu_a$ from gradients can be estimated as
 \be \label{ang}
 \mu_a(w) \, \sim \, 2 \pi \eta^2 \int_w^{r_A} \frac{1}{r} f(r)^2 \, .
 \ee
It then follows from (\ref{ang}) that 
 \be
 \frac{\partial}{\partial w} \mu_a(w) \, \sim \, - 2\pi \eta^2 \frac{1}{w} \, .
 \ee
 Hence, by minimizing the sum of potential and angular gradient energy (to first approximation the radial gradient energy does not depend on $w$) it follows that
 \be
 w \, \sim \, V_0^{-1/2} \eta \, .
 \ee
 For Abelian Higgs strings this yields
 \be
 w_{AH} \, \sim \lambda^{-1/2} \eta^{-1} \, ,
 \ee
 while for thermal inflation strings
 \be
 w \, \sim \, m_0^{-1} \, .
 \ee

 In terms of the phase transition temperature, the widths of thermal inflation strings and Abelian Higgs strings are of the same order of magnitude.  However, in terms of the mass per unit length there is a parametric difference, and this difference will have important implications for the string loop distribution.  From energy equipartition, the energy per unit length $\mu$ of a string is given by
 \be
 \mu \, \sim \, w^2 V_0 \, .
 \ee
 For Abelian Higgs strings this yields
 \be \label{massAH}
 \mu_{AH} \, \sim \, \eta^2 \, \sim \, T_c^2 \, ,
 \ee
 while for thermal inflation strings
 \be \label{massTI}
 \mu \, \sim \, \lambda^{-1/2} \frac{m_{pl}}{m_0} T_c^2 \, .
 \ee
 Thus,  for a fixed mass per unit length, a thermal inflation string has a width 
 \be
 w \, \sim \, \lambda^{-1/2} \frac{m_{pl}}{\sqrt{\mu}} \mu^{-1/2} \, ,
 \ee
 which is much greater than the width 
 \be
 w_{AH} \, \sim \, \lambda^{-1/2} \mu^{-1/2} \, .
 \ee
 of an Abelian Higgs string.
 
 Comparing (\ref{massAH}) and (\ref{massTI}), we see that for fixed symmetry breaking scale of $T_c \sim m_0 \sim 10^2 {\rm{GeV}}$, Abelian Higgs strings would have a mass per unit length of $G \mu_{AB} \,   \sim \, 10^{-34} $ which is many orders of magnitude smaller than the range of valiues of $G\mu$ which can have interesting cosmological effects. In the case of thermal inflation strings, on the other hand, for the same value of $T_c$ we obtain $G\mu \sim 10^{-17}$ which is now approaching the range which is of interest for string signals in cosmological observations.
 
 \section{Thermal String Loop Distribution}
 
 The parametric enhancement of the width of a thermal inflation string compared to the width of an Abelian Higgs string with the same mass per unit length has important implications for the loop distribution.
 
 The causality argument \cite{Kibble} which implies that the distribution of long strings (strings with a curvature radius comparable and greater than the Hubble radius) takes on a scaling solution where the number of long string segments crossing any given Hubble volume is independent of time applies equally to Abelian Higgs and thermal inflation strings. This {\it scaling solution} of the long string network is maintained by string loop production.  Like for Abelian Higgs strings, we can assume that the one scale loop production model \cite{onescale} also applies to thermal inflation strings, implying that at time $t$ loops are produced with radius $R = \alpha t$, where $\alpha$ is a constant which can be normalized by string evolution simulations which yield $\alpha \sim 10^{-1}$. Once produced, the number density $n(R, t)$ of loops in the radius interval between $R$ and $R + dR$ redshifts. Thus at times $t$ after the time $t_{eq}$ of equal matter and radiation we have
 \ba \label{dist}
 n(R, t) dR \, &=& \, N R^{-2} t^{-2} dR \,\,\,\, \alpha t > R > \alpha t_{eq}  \\
 n(R, t) dR \, &=& \, N R^{-5/2} t_{eq}^{1/2} t^{-2} dR \,\,\,\, \alpha t_{eq} > R > R_{co} \, ,
 \nonumber
 \ea
where $N$ is a constant determined by the number of long string segments per Hubble volume.  $R_{co}$ is a cutoff radius below which loops live for less than one Hubble time, and whose consequences for cosmology can be neglected.
 
For non-superconducting strings there are two main mechanisms by which string loops decay. The first is gravitational radiation: string loops have relativistic tension and hence oscillate and emit gravitational radiation. The power of gravitational radiation from a string loop of radius $R$ is \cite{Tanmay}
\be \label{GWdecay}
P_g \, = \gamma G \mu^2 \, ,
\ee
where $\gamma$ is a constant of the order $10^2$.  Gravitational radiation implies that loops with radius $R < R_g$ where
\be
R_g \, = \, \gamma G \mu t 
\ee
will live less than one Hubble expansion time,  and hence their cosmological effects are negligible.

Cusp evaporation is a second decay mechanism \cite{cuspdecay}.  A cusp is a point on the string which moves at the speed of light. Strings have finite width, and around the cusp point the string segments on either side of the cusp point overlap for a region of length \footnote{See \cite{Olum} for the correction of an error present in the original work of \cite{cuspdecay}.} 
\be
l_c(R) \, \sim \, R^{1/2} w^{1/2} \, .
\ee
Locally the cusp region looks like a string-antistring pair, and there is no topology protecting the cusp region against annihilation into gauge and scalar field quanta.  It can be proven that string loops described by the effective Nambu-Goto action have at least one cusp per oscillation time \cite{KT}.  Hence, the power of the cusp annihilation process is
\be \label{cuspdecay}
P_c \, \sim \, \frac{1}{R} l_c(R) \mu \, = \, \bigl( \frac{w}{R} \bigr)^{1/2} \mu \, .
\ee
From the above equation it follows that, due to cusp evaporation, string loops with radius less than
\be
R_c \, = \, w^{1/3} t^{2/3}
\ee
live for less than one Hubble expansion time.  The cutoff radius $R_{co}$ in the loop distribution of (\ref{dist}) is the larger of $R_g$ and $R_c$.

Comparing the strengths of gravitational radiation power (\ref{GWdecay}) and cusp annihilation power (\ref{cuspdecay}) we see that the parametrically larger width of a thermal inflation string (for a given mass per unit length) will lead to a parametric amplification of the role of cusp annihilation compared to gravitational wave decay.  We also see that the relative importance of cusp annihilation increases the lower the value of $G\mu$ is.

The condition for cusp annihilation to dominate over gravitational radiation is $R_c > R_g$ or
\be
w \, > \, (\gamma G \mu)^3 t \, .
\ee
For Abelian Higgs strings this yields
\be
\bigl( \frac{T}{m_{pl}} \bigr)^2 \, > \, \lambda^{1/2} \gamma^3 (G \mu)^{7/2} \, ,
\ee
or, expressed in terms of the critical temperature $T_c$
\be \label{AHcrit}
\frac{T}{T_c} \, > \, \lambda^{1/4} \gamma^{3/2} \bigl(  \frac{T}{m_{pl}} \bigr)^{5/2} \, .
\ee
On the other hand, for thermal inflation strings we obtain
\be
\bigl( \frac{T}{m_{pl}} \bigr)^2 \, > \, \gamma^3 (G \mu)^4 \, ,
\ee
or,  after expressing $\mu$ in terms of the critical temperature
\be \label{TIcrit}
\frac{T}{T_c} \, > \, \gamma^{3/2} \lambda^{-1} \frac{T_c}{m_{pl}} \, .
\ee

Comparing these expressions we see that, for a fixed value of the string tension, cusp annihilation is more important for thermal inflation strings than for Abelian Higgs strings. On the other hand, fixing $T_c$ we see that the importance of cusp annihilation is, maybe surprisingly, less than for Abelian Higgs strings.

Evaluating (\ref{AHcrit}) and (\ref{TIcrit}) at the temperature $T_{eq} \sim 1 {\rm{eV}}$ of equal matter and radiation (the temperature relevant for cosmological signatures of strings), we see that for Abelian Higgs strings the cutoff in the loop distribution is determined by cusp annihilation for values $T_c <  10^{10} {\rm{GeV}}$ while for thermal inflation strings it is for values $T_c < 10^8 {\rm{GeV}}$. In particular, for the value $m_0 \sim 10^2 - 10^3 {\rm{GeV}}$ assumed in the original thermal inflation paper \cite{Lyth96}, we conclude that the cutoff in the loop distribution is given by the cusp annihilation process.

\section{Constraints from Cosmological Observations}

In this section we study what constraints on the symmetry breaking scale $m_0$ of thermal inflation can be derived from cosmological observations. Cosmic strings leave behind interesting signals in many observational windows.  In most cases, the effects are gravitational, and hence the magnitude of the string signal depends on $G\mu$.  In terms of $G\mu$, the strength of the signals will hence be the same for Abelian Higgs and thermal inflation strings. However, since the relation between $T_c$ and $\mu$ is different, the magnitude of the string signals as a function of $T_c$ will change.  As the string network consists of both long strings and loops, each will induce specific signatures. 

We first turn to signatures of the long string network.  For example, long strings lead to line discontinuities in cosmic microwave background (CMB) anisotropy maps \cite{KS}.    This is due to the fact that a long straight string produces a conical deformation of the metric with deficit angle proportional to $G\mu$ \cite{deficit}.The magnitude of this signal depends on $G\mu$. The study of \cite{Hergt} shows that experiments with the specifications of the South Pole Telescope or the Atacama Cosmology Telescope can constrain the string tension to be
\be \label{bound1}
G\mu \, < \, 10^{-8} \, .
\ee
A slighty weaker bound of
\be
G\mu \, < \, 10^{-7} 
\ee
can be derived from the angular power spectrum of CMB anisotropies \cite{Dvorkin}. The resulting bound on $T_c$ for thermal inflation strings is parametrically stronger than for Abelian Higgs strings, namely
\be
\frac{T_c}{m_{pl}} \, < \, \lambda^{1/2} 10^{-7} \, .
\ee
This bound is obviously satisfied for the value $T_c \sim m_0 = 10^2 - 10^3 {\rm{GeV}}$ assumed in \cite{Lyth96}.

For Abelian Higgs strings there is a tighter bound of
\be
G\mu \, < \, 10^{-10} 
\ee
which comes from the upper bound on the stochastic background of gravitational waves from pulsar timing array measurements \cite{PTA}.  This bound depends on having a scaling distribution of loops down to the gravitational radiation cutoff $R_g$.  This bound remains valid for thermal inflation strings since, as the discussion at the end of the previous section showed, cusp annihilation only changes the loop distribution for values of $G\mu$ which are lower than the above bound.

Long strings moving through space produce overdense regions in their wake \cite{wake}. CMB photons passing through these wakes get absorbed at the 21-cm wavelength. Long cosmic strings hence lead to distinct signals in high redshift 21-cm surveys \cite{Holder2}: wedges of absorption in 21-cm redshift maps which are extended in the angular directions and narrow in redshift direction.  The study of \cite{Maibach} shows that the string signal can be detected by surveys such as the MWA telescope down to a value of $G\mu$ comparable to that of (\ref{bound1}), and prospects indicate that with better analysis tools a significant improvement of this bound can be expected. 

String wakes also lead to rectangles in the sky with induced CMB polarization (including a B-mode component) \cite{Holder1}. From the analysis of \cite{Hannah} it appears, however, that this signal is harder to extract from observations than the 21-cm signal.  At lower redshifts, string wakes also lead to planar overdensities of galaxies.  These are, however, disrupted by the gravitational effects of the dominant source of fluctuations \cite{Disrael}.

Turning now to signatures of cosmic string loops, the gravitational signals are the same for Abelian Higgs strings and thermal inflation strings given the same value of $G\mu$.  String loops seed nonlinear structures by gravitational accretion \cite{early}. Given the bounds on $G\mu$ discussed above,  strings can only play a subdominant role in explaining the nonlinear structures today. However, since strings form nonlinear seeds immediately after their formation, they will dominate the halo mass function at sufficiently early times \cite{Jiao}. They can provide seeds for intermediate and super-massive black holes at high redshifts \cite{Jerome}. It has recently been shown \cite{Bryce} that for superconducting cosmic strings the ``Direct Collapse Black Hole'' criteria can be satisifed, and that such loops indeed could explain the origin of the observed high redshift super-massive black holes.

Since thermal inflation strings have a greater width than Abelian Higgs strings for a fixed value of $T_c$, non-gravitational signals from thermal inflation strings may differ from those of Abelian Higgs strings. Specifically, the flux of cosmic rays \cite{CR} due to cosmic strings will be larger \cite{prep}.

\section{Conclusions and Discussion} \label{conclusion}

 We have pointed out that thermal inflation models lead to the production of a network of cosmic strings. These thermal inflation strings have different poperties compared to strings arising in the Abelian Higgs model. Specifically, for a fixed phase transition temperature,  thermal inflation strings have a larger mass per unit length, and hence lead to larger gravitational effects.
 
For thermal inflation strings, the upper bound on the phase transition temperature from cosmological observations are parametrically more stringent than for Abelian Higgs strings. However, for the value of the symmetry breaking scale suggested in \cite{Lyth96}, the bounds are satisfied.
 
\section*{Acknowledgement}

RB wishes to thank the Pauli Center and the Institutes of Theoretical Physics and of Particle- and Astrophysics of the ETH for hospitality. The research at McGill is supported in part by funds from NSERC and from the Canada Research Chair program.

\end{document}